\documentclass[runningheads]{llncs}
\usepackage[T1]{fontenc}
\usepackage{amsfonts,amssymb}
\usepackage{multirow}
\usepackage{amsmath}
\usepackage{graphicx}
\usepackage{subcaption}
\usepackage{mathrsfs}
\usepackage{caption}
\usepackage{amssymb} 
\usepackage{booktabs}
\usepackage{hyperref}
\usepackage{colortbl}
\usepackage{makeidx}
\usepackage{tabularx}
\makeindex
\DeclareCaptionLabelSeparator{dot}{. }
\usepackage{siunitx} 
\usepackage[table]{xcolor} 
\usepackage{subcaption}
\usepackage{xcolor}
\usepackage[misc]{ifsym}
\usepackage{booktabs}
\hypersetup{
  colorlinks=true,
  linkcolor=blue,
  citecolor=blue, 
  urlcolor=blue, 
  linkbordercolor=white, 
  citebordercolor=white
}
\definecolor{lightyellow}{rgb}{1.0, 1.0, 0.6}
\begin{document}
\title{Centerline Boundary Dice Loss for Vascular Segmentation}
\author{
  Pengcheng Shi\inst{1} 
  \and Jiesi Hu\inst{1,2}
  \and Yanwu Yang\inst{1,2}
  \and \\ Zilve Gao\inst{1}
  \and Wei Liu\inst{1}
  \and Ting Ma\inst{1,2,3,4}$^{(\textrm{\Letter})}$
}

\authorrunning{P. Shi et al.}

\institute{Electronic \& Information Engineering School, Harbin Institute of Technology (Shenzhen), Shenzhen, China\\ \email{tma@hit.edu.cn} \and
Peng Cheng Laboratory, Shenzhen, China \and
Guangdong Provincial Key Laboratory of Aerospace Communication and Networking Technology, Harbin Institute of Technology (Shenzhen), Shenzhen, China \and International Research Institute for Artificial Intelligence, Harbin Institute of Technology (Shenzhen), Shenzhen, China}

\maketitle              
\begin{abstract} 
Vascular segmentation in medical imaging plays a crucial role in analysing morphological and functional assessments. Traditional methods, like the centerline Dice (clDice) loss, ensure topology preservation but falter in capturing geometric details, especially under translation and deformation. The combination of clDice with traditional Dice loss can lead to diameter imbalance, favoring larger vessels. Addressing these challenges, we introduce the centerline boundary Dice (cbDice) loss function, which harmonizes topological integrity and geometric nuances, ensuring consistent segmentation across various vessel sizes. cbDice enriches the clDice approach by including boundary-aware aspects, thereby improving geometric detail recognition. It matches the performance of the boundary difference over union (B-DoU) loss through a mask-distance-based approach, enhancing traslation sensitivity. Crucially, cbDice incorporates radius information from vascular skeletons, enabling uniform adaptation to vascular diameter changes and maintaining balance in branch growth and fracture impacts. Furthermore, we conducted a theoretical analysis of clDice variants (cl-X-Dice). We validated cbDice's efficacy on three diverse vascular segmentation datasets, encompassing both 2D and 3D, and binary and multi-class segmentation. Particularly, the method integrated with cbDice demonstrated outstanding performance on the MICCAI 2023 TopCoW Challenge dataset. Our code is made publicly available at: \url{https://github.com/PengchengShi1220/cbDice}.
\end{abstract}
\section{Introduction}
\begin{figure}[t]
\centering
\includegraphics[width=\textwidth]{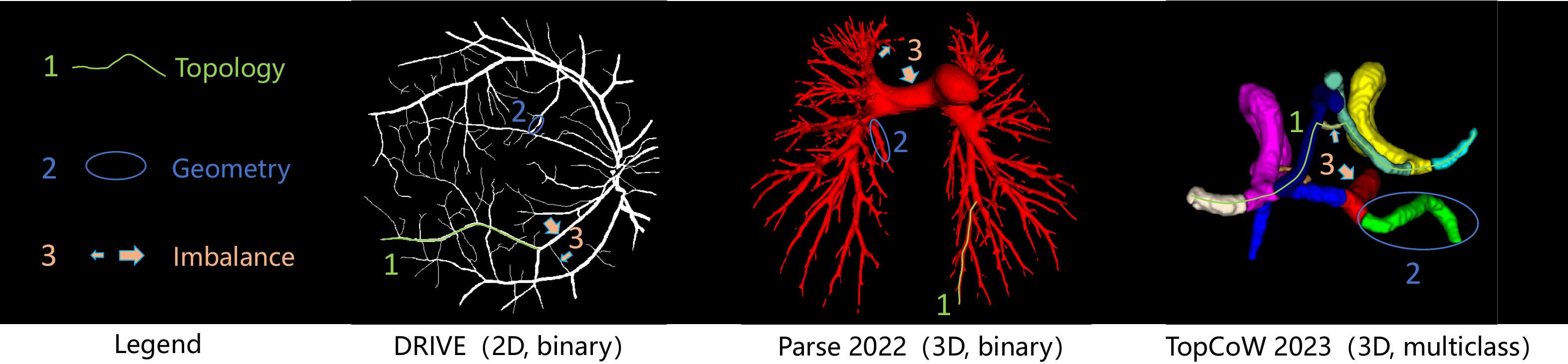}
\caption{Challenges in vascular segmentation across diverse datasets. Key features are highlighted: (1) green centerlines for vessel connectivity; (2) blue circles for morphological characteristics; and (3) light orange arrows within light blue frames indicating branches with significant diameter differences.} \label{fig1}
\end{figure}

Advancements in deep learning have remarkably enhanced medical image segmentation, especially in vasculature, revolutionizing diagnostic and interventional radiology. However, prevalent metrics frequently neglect domain-specific needs, exposing significant deficiencies \cite{maier2024metrics,reinke2024understanding}. Vascular segmentation encounters three significant challenges: (1) preserving the topology of the vascular network to enable accurate hemodynamic analysis; (2) capturing the intricate geometric morphologies vital for the diagnosis of conditions like stenosis; (3) attaining balanced segmentation across vessel diameters and consistent width adjustments within each branch to mitigate diameter imbalance (see Fig.~\ref{fig1}). \\
\indent{In this paper, we seek to address a fundamental question: \textit{How to encompass topology, geometry, and vessel diameter consistency in vascular segmentation?} Recent advancements in the field of vascular segmentation have been significantly driven by the integration of deep neural networks that utilize both topological \cite{hu2019topology,qi2023dynamic,stucki2023topologically,shit2021cldice,menten2023skeletonization,li2023robust,qiu2023corsegrec,gupta2024topology} and geometric \cite{xia20223d,qi2023dynamic,wang2020deep,nikolov2021clinically,song2022segmentation,qiu2023corsegrec} principles. While significant strides have been made, particularly in designing loss functions to mitigate data imbalance in medical image segmentation \cite{kervadec2019boundary,zhao2020rethinking,taghanaki2019combo,yeung2022unified}, a comprehensive analytical framework in this domain remains elusive. Our study aims to bridge this gap by proposing the centerline boundary Dice (\textbf{cbDice}) loss function. Distinctively, cbDice integrates boundary awareness with radius information extracted from vascular skeletons. This methodology facilitates equitable segmentation across vessels of differing diameters, critically preserving topological integrity. Our contributions in this study are multifaceted: \textbf{(1)} We introduce the unified clDice variants (cl-X-Dice) loss, a new framework combining topology, geometry, and vessel diameter consistency for vascular segmentation. \textbf{(2)} We refine the centerline Dice (clDice) loss with geometric considerations, equating its performance to the boundary difference over union (B-DoU) loss through a mask-distance-based method that enhances traslation sensitivity. \textbf{(3)} We enhance the cbDice loss by integrating radius information from vascular skeletons. This enables uniform adaptation to changes in vascular diameter, ensuring a balanced impact on fractures or growth across various branches. \textbf{(4)} We conduct a theoretical analysis to show the scrutinizing their responsiveness to geometric transformations among different cl-X-Dice implementations. Additionally, we evaluate cbDice loss across a range of vascular segmentation datasets, covering 2D and 3D dimensions and both binary and multi-class segmentation tasks.} \\

\noindent\textbf{Related works:} \textbf{(i) Topologically}, maintaining connectivity, as underscored by persistent homology \cite{hu2019topology,clough2020topological,qi2023dynamic,stucki2023topologically}, centerline or skeletonization techniques \cite{shit2021cldice,menten2023skeletonization}, and Euler characteristic methods \cite{li2023robust}, is essential. The emergence of topology-aware uncertainty estimation \cite{gupta2024topology} has also played a significant role. Despite advancements, challenges remain. For example, the clDice loss function \cite{shit2021cldice}, though effective in ensuring topological integrity, has limitations in capturing geometric nuances, especially during translation and deformation. This underscores the need for approaches that harmoniously balance topological fidelity with precise morphological representation in vessel imagery. \textbf{(ii) Geometrically}, boundary precision is highlighted by methods such as the B-DoU loss \cite{sun2023boundary} and edge-reinforced networks \cite{xia20223d}. Techniques such as dynamic snake convolution \cite{qi2023dynamic} target fine, intricate features. In medical image segmentation, distance map-based approaches are crucial, including shape-aware segmentation with signed distance maps \cite{xue2020shape}, and CNN integration of distance transform maps \cite{ma2020distance}. Recent techniques that combine centerlines with distance maps have demonstrated the capability to constrain both the skeleton and the geometric morphology in tubular structure segmentation \cite{wang2020deep,song2022segmentation}. However, they lack comprehensive integration within a loss function, a crucial aspect for enhancing applicability and generalization, and have not been adequately linked with centerline-based metrics like the clDice. Although the normalized skeleton distance transform (NSDT) clDice loss \cite{qiu2023corsegrec} considers geometry based on clDice, this method overlooks the challenges posed by diameter imbalance. \textbf{(iii) Imbalance} is a common issue in medical image segmentation, where traditional loss functions like the Dice loss exhibit a bias towards larger anatomical structures. To address this issue, recent studies have introduced a range of targeted solutions, including boundary loss \cite{kervadec2019boundary}, focal Dice loss \cite{zhao2020rethinking}, combo loss \cite{taghanaki2019combo}, and unified focal loss \cite{yeung2022unified}. These approaches are designed to mitigate imbalanced segmentation. However, they do not specifically address the problem of imbalanced vessel diameter in vascular segmentation, while also neglecting the connectivity of the vascular network.\\

\section{Methodology}
\noindent\textbf{Preliminaries.} This method first processes an input \( X \in \mathbb{R}^{c_{\mathrm{i}} \times \mathcal{N}} \) through a model to generate an output \( Y \in \mathbb{R}^{c_{\mathrm{o}} \times \mathcal{N}} \), where \( c_{\mathrm{i}} \) and \( c_{\mathrm{o}} \) denote the number of input and output channels, respectively. Subsequently, the output \( Y \) is transformed into a binary mask, denoted by \( V \). Here, \( \mathcal{N} \) signifies the total number of pixels or voxels, defined as \( w \times h \) for 2D images, or \( w \times h \times d \) for 3D volumes, where \( w \), \( h \), and \( d \) represent width, height, and depth, respectively. The mask \( V \) is defined as \( V = \{v_i, b^{\text{v}}_j \mid i \in [1, q], j \in [1, k] \} \), comprising \( q \) mask points (\( v \)) with value 1, and \( k \) background points (\( b^{\text{v}} \)) with value 0. The corresponding skeleton \( S \), derived from \( V \), consisting of \( S = \{s_i, b^{\text{s}}_j \mid i \in [1, n], j \in [1, m] \} \) with \( n \) skeletal points (\( s \)) valued 1, and \( m \) background points (\( b^{\text{s}} \)) valued 0. Both \( V \) and \( S \) belong to \( \mathbb{R}^{\mathcal{N}} \). The subscripts \text{P} and \text{L} respectively denote the prediction and the reference. We begin by examining the traditional clDice\cite{shit2021cldice}, introduced for topology preservation:
\begin{align}
\mathrm{Tprec}(S_{\mathrm{P}}, V_{\mathrm{L}}) = \frac{|S_{\mathrm{P}} \cap V_{\mathrm{L}}|}{|S_{\mathrm{P}}|}, \quad \mathrm{Tsens}(S_{\mathrm{L}}, V_{\mathrm{P}}) = \frac{|S_{\mathrm{L}} \cap V_{\mathrm{P}}|}{|S_{\mathrm{L}}|}
\end{align}
\begin{equation}
\mathrm{clDice}(V_{\mathrm{P}}, V_{\mathrm{L}}) = \frac{2 \times \mathrm{Tprec}(S_{\mathrm{P}}, V_{\mathrm{L}}) \times \mathrm{Tsens}(S_{\mathrm{L}}, V_{\mathrm{P}})}{\mathrm{Tprec}(S_{\mathrm{P}}, V_{\mathrm{L}}) + \mathrm{Tsens}(S_{\mathrm{L}}, V_{\mathrm{P}})}
\end{equation}
where \( V_{\mathrm{P}} \) and \( V_{\mathrm{L}} \) are the masks of predicted and labeled segments. The skeletons \( S_{\mathrm{P}} \) and \( S_{\mathrm{L}} \) are extracted from \( V_{\mathrm{P}} \) and \( V_{\mathrm{L}} \) respectively. \\
\begin{figure}[t]
\centering
\includegraphics[width=\textwidth]{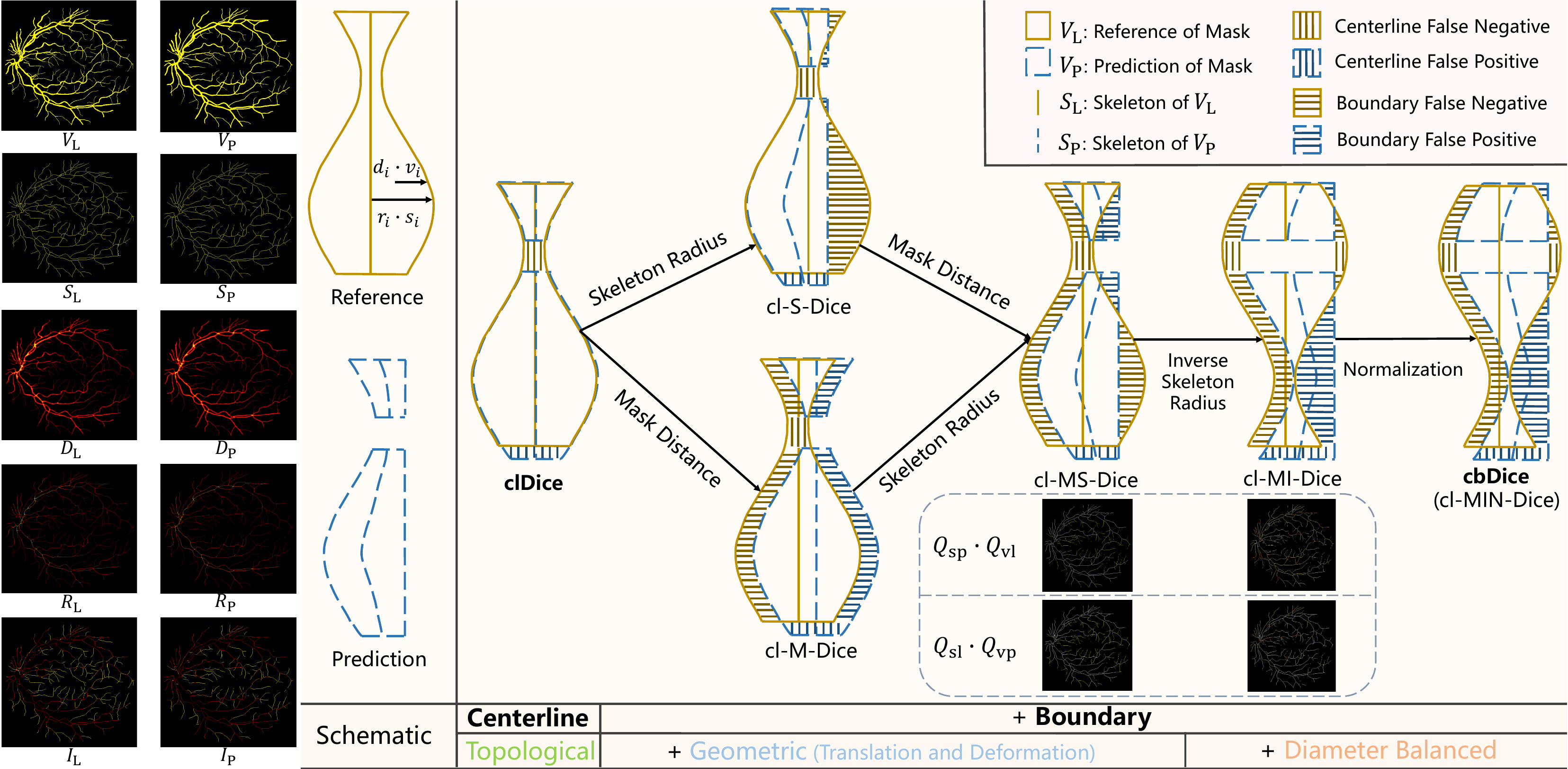}
\caption{This figure displays a segmented 2D retinal vessel, offering a visual example of the different variables mentioned in Table~\ref{tab:comprehensive_metrics_comparison}. A 2D schematic depicts the transition from clDice to cbDice, visualized as a bottle with varying radius.} \label{fig2}
\end{figure}
\noindent\textbf{Variations of clDice.} Fig.~\ref{fig2} illustrates the stepwise evolution of the cl-X-Dice. Our method models tubular structures as entities with variable radius: employing normals to the skeleton line for depicting radius changes in 2D, and using circular cross-sections aligned along the central line in 3D. This alignment facilitates volume estimation through the application of set-based expressions. The set \( D \) signifies the minimum distances from \( q \) mask points (\( v \)) in \( V \) to their respective boundaries, defined as \( D = \{d_i \cdot v_i, b^{\text{v}}_j \mid i \in [1, q], j \in [1, k] \} \), thus forming the distance map. The set \( R \), associated with skeletal points in \( S \), is defined as \( R = \{r_i \cdot s_i, b^{\text{s}}_j \mid i \in [1, n], j \in [1, m] \} \), representing \( n \) skeletal points (\( s \)) with corresponding skeletal radius (\( r \)). The radius of each skeletal point in \( S \) is derived from the distance map \( D \), resulting in the set \( R \). Conversely, \( I = \{\frac{s_i}{r_i}, b^{\text{s}}_j \mid i \in [1, n], j \in [1, m] \} \) denotes the set of inverse skeleton radius for the same \( n \) skeletal points in \( S \). \( D \), \( R \) and \( I \) are elements of \( \mathbb{R}^{\mathcal{N}} \). To reconcile differences between skeleton line and distance map computations, values in \( D \) exceeding \( R_{\max} \) are adjusted to \( R_{\max} \). Hence, the range of \( R \) is [0, \( R_{\max} \)] and the range of \( D \) is [1, \( R_{\max} \)]. For the \( p \) mask points \( \{v_i \mid i \in \mathbb{Z}, i \in [1, p] \} \) located either in a 2D plane perpendicular to the centerline or in a 3D cross-section, which include a skeleton point \( s \) with a corresponding skeleton radius \( r \), the distance \( d_i \) from each of the \( p \) mask points to the boundary is within the range of [1, \( r \)]. The \text{cl-X-Dice} metric is designed to address segmentation challenges in vasculatures of varying diameters:
\begin{equation}
\label{eq:Tprec_clXDice}
\mathrm{Tprec}(S_{\mathrm{P}}, S_{\mathrm{L}}, V_{\mathrm{L}}) = \frac{|Q_{\mathrm{sp}} \cap Q_{\mathrm{vl}}|}{|Q_{\mathrm{sp}} \cap Q_{\mathrm{spvp}} \cap (U - S_{\mathrm{L}})| + |Q_{\mathrm{sp}} \cap Q_{\mathrm{slvl}}|} \\
\end{equation}
\begin{equation}
\label{eq:Tsens_clXDice}
\mathrm{Tsens}(S_{\mathrm{L}}, S_{\mathrm{P}}, V_{\mathrm{P}}) = \frac{|Q_{\mathrm{sl}} \cap Q_{\mathrm{vp}}|}{|Q_{\mathrm{sl}} \cap Q_{\mathrm{slvl}} \cap (U - S_{\mathrm{P}})| + |Q_{\mathrm{sl}} \cap Q_{\mathrm{spvp}}|} \\
\end{equation}
\begin{equation}
\label{eq:cl-X-Dice}
\text{cl-X-Dice}(V_{\mathrm{P}}, V_{\mathrm{L}}) = \frac{2 \times \mathrm{Tprec}(S_{\mathrm{P}}, S_{\mathrm{L}}, V_{\mathrm{L}}) \times \mathrm{Tsens}(S_{\mathrm{L}}, S_{\mathrm{P}}, V_{\mathrm{P}})}{\mathrm{Tprec}(S_{\mathrm{P}}, S_{\mathrm{L}}, V_{\mathrm{L}}) + \mathrm{Tsens}(S_{\mathrm{L}}, S_{\mathrm{P}}, V_{\mathrm{P}})}
\end{equation}
where we define \( U \in \mathbb{R}^{\mathcal{N}} \) with each element set to 1. We introduce variables \( Q_{\mathrm{sp}}, Q_{\mathrm{sl}}, Q_{\mathrm{vl}}, \) and \( Q_{\mathrm{vp}} \) to address geometric and vascular diameter imbalance aspects in segmentation processes. These variables are crucial for the detailed analysis of vascular structures. A comprehensive comparison of the \text{cl-X-Dice} for vascular segmentation is available in Table~\ref{tab:comprehensive_metrics_comparison}. Key notations for \( Q \) are defined as: \( \mathrm{sp} = S_{\mathrm{P}}, \mathrm{sl} = S_{\mathrm{L}}, \mathrm{vl} = V_{\mathrm{L}}, \mathrm{vp} = V_{\mathrm{P}}, \mathrm{slvl} = S_{\mathrm{L}} \cap V_{\mathrm{L}}, \) and \( \mathrm{spvp} = S_{\mathrm{P}} \cap V_{\mathrm{P}} \). Additionally, normalized ratios are represented as \( R_{\text{N}} = \frac{R}{R_{\max}}, I_{\text{N}} = \frac{I}{I_{\min}}, \) and \( D_{\text{N}} = \frac{D}{R_{\max}} \). Figure \ref{fig3} demonstrates how different metrics respond to translation, deformation, and diameter imbalance. We extensively analyze the cl-X-Dice metric's theoretical response to geometric transformations. For a detailed proof, see the supplementary materials.\\
\begin{table}[t]
\centering
\caption{Stepwise evolution of cl-X-Dice and comparison of 2D and 3D metrics. The abbreviations include reference (\(\mathrm{L}\)), prediction (\(\mathrm{P}\)), centerline (\(\mathrm{cl}\)), Dice (\(\mathrm{D}\)), skeleton (\(\mathrm{S}\)), mask (\(\mathrm{M}\)), inverse skeleton radius (\(\mathrm{I}\)), normalized (\(\mathrm{N}\)), and centerline boundary (\(\mathrm{cb}\)). \textbf{cl-MIN-D} is equivalent to \textbf{cb-D} in this study.}
\label{tab:comprehensive_metrics_comparison}
\renewcommand{\arraystretch}{1} 
\begin{tabular*}{\textwidth}{@{} l @{\extracolsep{\fill}} *{14}{S[table-format=-1.2]} @{}} 
\toprule
{Metric} & \multicolumn{2}{c}{cl-D} & \multicolumn{2}{c}{cl-S-D} & \multicolumn{2}{c}{cl-M-D} & \multicolumn{2}{c}{cl-MS-D} & \multicolumn{2}{c}{cl-MI-D} & \multicolumn{2}{c}{cl-MSN-D} & \multicolumn{2}{c}{\textbf{cl-MIN-D}} \\
\cmidrule(lr){2-3} \cmidrule(lr){4-5} \cmidrule(lr){6-7} \cmidrule(lr){8-9} \cmidrule(lr){10-11} \cmidrule(lr){12-13} \cmidrule(lr){14-15}
{Dim} & {2D} & {3D} & {2D} & {3D} & {2D} & {3D} & {2D} & {3D} & {2D} & {3D} & {2D} & {3D} & {2D} & {3D} \\
\midrule
\( Q_{\mathrm{sl}} \) & {\( S_{\mathrm{L}} \)} & {\( S_{\mathrm{L}} \)} & {\( R_{\mathrm{L}} \)} & {\( R^2_{\mathrm{L}} \)} & {\( S_{\mathrm{L}} \)} & {\( S_{\mathrm{L}} \)} & {\( R_{\mathrm{L}} \)} & {\( R^2_{\mathrm{L}} \)} & {\( I_{\mathrm{L}} \)} & {\( I^2_{\mathrm{L}} \)} & { \( R_{\mathrm{L}, \text{N}} \)} & {\( R^2_{\mathrm{L}, \text{N}} \)} & {\( I_{\mathrm{L}, \text{N}} \)} & {\( I^2_{\mathrm{L}, \text{N}} \)} \\
\( Q_{\mathrm{sp}} \) & {\( S_{\mathrm{P}} \)} & {\( S_{\mathrm{P}} \)} & {\( R_{\mathrm{P}} \)} & {\( R^2_{\mathrm{P}} \)} & {\( S_{\mathrm{P}} \)} & {\( S_{\mathrm{P}} \)} & {\( R_{\mathrm{P}} \)} & {\( R^2_{\mathrm{P}} \)} & {\( I_{\mathrm{P}} \)} & {\( I^2_{\mathrm{P}} \)} & {\( R_{\mathrm{P}, \text{N}} \)} & {\( R^2_{\mathrm{P}, \text{N}} \)} & {\( I_{\mathrm{P}, \text{N}} \)} & {\( I^2_{\mathrm{P}, \text{N}} \)} \\
\cmidrule(lr){2-5} \cmidrule(lr){6-11} \cmidrule(lr){12-15}
\( Q_{\mathrm{vl}} \) & \multicolumn{4}{c}{\( V_{\mathrm{L}} \)} & \multicolumn{6}{c}{\( D_{\mathrm{L}} \)} & \multicolumn{4}{c}{\( D_{\mathrm{L}, \text{N}} \)} \\
\( Q_{\mathrm{vp}} \) & \multicolumn{4}{c}{\( V_{\mathrm{P}} \)} & \multicolumn{6}{c}{\( D_{\mathrm{P}} \)}  & \multicolumn{4}{c}{\( D_{\mathrm{P}, \text{N}} \)} \\
\( Q_{\mathrm{slvl}} \) & \multicolumn{4}{c}{\( S_{\mathrm{L}} \)} & \multicolumn{6}{c}{\( R_{\mathrm{L}} \)} & \multicolumn{4}{c}{\( R_{\mathrm{L}, \text{N}} \)} \\
\( Q_{\mathrm{spvp}} \) & \multicolumn{4}{c}{\( S_{\mathrm{P}} \)} & \multicolumn{6}{c}{\( R_{\mathrm{P}} \)} & \multicolumn{4}{c}{\( R_{\mathrm{P}, \text{N}} \)} \\
\bottomrule
\end{tabular*}
\end{table}

\section{Experiments}
\subsection{Datasets}
We evaluate our proposed methodology across three vascular segmentation datasets. The DRIVE dataset \cite{staal2004ridge}, a benchmark for 2D retinal vessel imaging, includes 40 color fundus photographs (584x565 pixels), split into 16 training, 4 validation, and 20 testing images. For 3D CT pulmonary artery segmentation, the MICCAI 2022 Parse Challenge dataset \cite{luo2023efficient} comprises 100 CT scans (512x512x228 to 512x512x376 pixels), with 80 for training and 20 for validation and testing. Lastly, the MICCAI 2023 TopCoW Challenge dataset \cite{yang2023benchmarking} consists of 90 brain MRA cases (approximately 481x586x185 pixels), with 72 for training and 18 for validation and testing.
\begin{figure}[t]
\centering
\includegraphics[width=\textwidth]{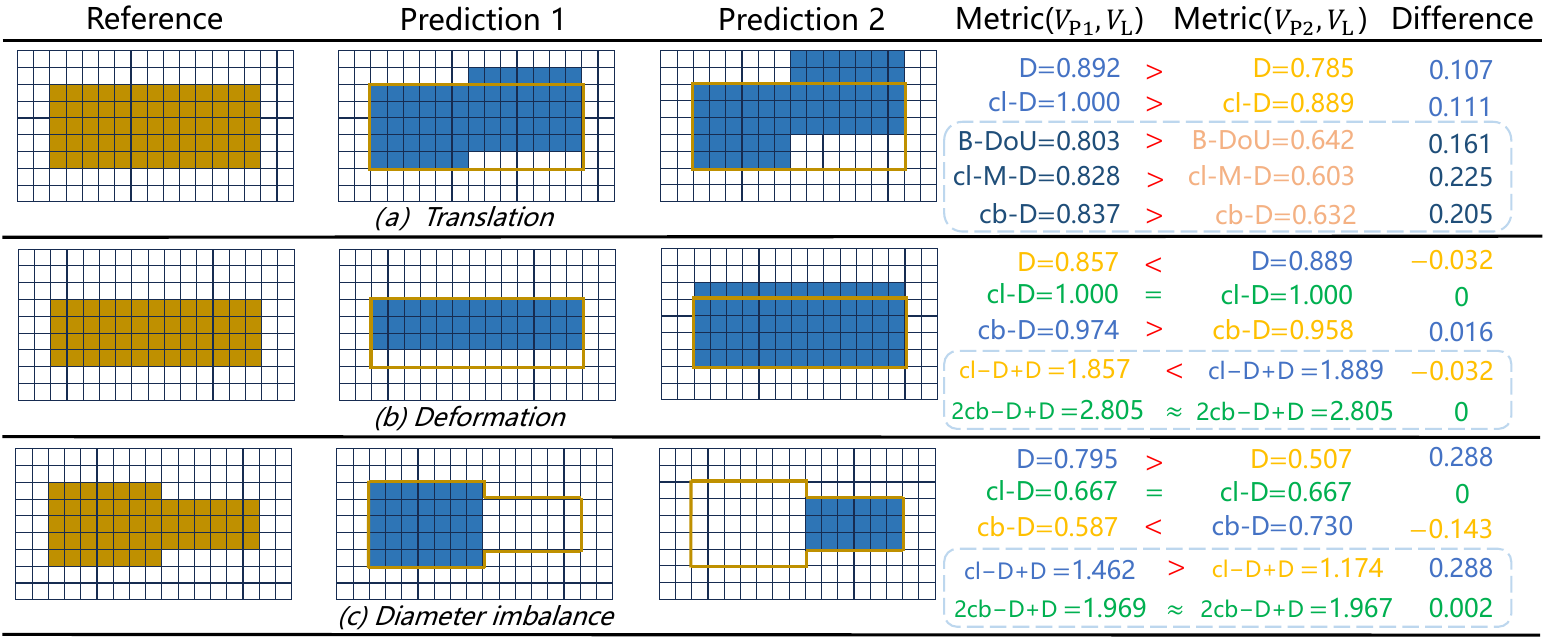}
\caption{(a) With translation-only perturbations, cb-Dice metric sensitivity to cl-M-Dice variations, increasing alongside translation distance, is comparable to B-DoU, while clDice remains near 1. (b) In uniform scaling (enlargement or reduction), cbDice-Dice pairing ensures more consistent evaluations than clDice-Dice, effectively adapting to scale changes. (c) For diameter imbalances, cbDice-Dice consistently assesses varied diameter branches, outperforming clDice-Dice.} \label{fig3}
\end{figure}
\begin{table}[t]
\centering
\caption{Comparison of DRIVE dataset results via nnU-Net over 20 epochs. Our methods are highlighted in light yellow, with best scores in \textbf{bold}.}
\label{tab:comprehensive_results_comparison_2D}
\begin{tabular*}{\textwidth}{@{\extracolsep{\fill}} c|c|ccccccccc}
\hline
\multirow{3}{*}{Loss} & \(\text{X}\) & -- & -- & B-DoU & B-DoU & \cellcolor{lightyellow} cl-M-Dice & clDice & \cellcolor{lightyellow} cbDice & \cellcolor{lightyellow} cbDice & \cellcolor{lightyellow} cbDice \\
                      & \( \alpha \) & 0 & 1 & 0 & 1 & \cellcolor{lightyellow}1 & 1 & \cellcolor{lightyellow}1 & \cellcolor{lightyellow}1 & \cellcolor{lightyellow}1 \\
                      & \( \beta \)  & 0 & 0 & 1 & 1 & \cellcolor{lightyellow}1 & 1 & \cellcolor{lightyellow}0.5 & \cellcolor{lightyellow}1 & \cellcolor{lightyellow}2 \\
\hline
\multirow{2}{*}{Overlap \(\uparrow\)} & Dice & 81.8 & 82.4 & 82.2 & 82.3 & 82.4 & 82.3 & \textbf{82.5} & 82.4 & 82.3 \\
                                      & clDice & 81.2 & 82.3 & 81.9 & 82.1 & 82.3 & 82.2 & 82.3 & \textbf{82.4} & 82.2 \\
\hline
\multirow{2}{*}{Topology \(\downarrow\)} & \(\beta^{\text{err}}\) & 345 & 360 & \textbf{339} & 343 & 356 & 355 & 358 & 361 & 351 \\
                                          & \(\mu^{\text{err}}\) & 631 & 599 & 620 & 611 & 599 & 605 & \textbf{590} & 601 & 607 \\
\hline
Distance \(\uparrow\) & NSD & 86.1 & 86.9 & 87.0 & 87.1 & 87.1 & 86.8 & \textbf{87.2} & 87.0 & 86.9 \\
\hline
\end{tabular*}
\end{table}
\subsection{Setup}
We conducted our experiments using PyTorch 2.1, utilizing NVIDIA V100 for increased computational efficiency. All experiments were trained from scratch on the nnU-Net V2 framework \cite{isensee2021nnu}. In standardizing our experimental setup, we opted not to use deep supervision, set the batch size to 2, and adjusted other parameters according to the default configurations of nnU-Net for each dataset. Additionally, for the TopCoW 2023 dataset, we disabled mirror augmentation to prevent the incorrect flipping of left and right labels. We rigorously assessed the effectiveness of the cbDice loss across a range of state-of-the-art segmentation models from the TopCoW 2023 Challenge \cite{yang2023benchmarking} and from other recent studies. This assessment included nnU-Net \cite{isensee2021nnu}, SwinUNETR \cite{hatamizadeh2021swin} and NexToU \cite{shi2023nextou}. The comparative study involved several loss functions: standard Dice loss, clDice loss \cite{shit2021cldice}, B-DoU loss \cite{sun2023boundary}, and our proposed cl-M-Dice and cbDice loss. Segmentation performance was evaluated using key metrics in overlap (Dice and clDice \cite{shit2021cldice}), topology (Betti Number error and Betti Matching error \cite{stucki2023topologically}), and distance (Normalized Surface Distance (NSD) at 1.0mm tolerance, following \cite{nikolov2021clinically,maier2024metrics}). \(\beta^{\text{err}}\) and \(\mu^{\text{err}}\) correspond to the Betti number error and the Betti matching error, respectively. We adopted the topology preserving differentiable skeleton extraction algorithm described in \cite{menten2023skeletonization}, and implemented the Euclidean distance transform using a GPU-accelerated approach with the cuCIM library. Aligning with nnU-Net, we maintain the Cross Entropy (CE) loss. The loss function is \(\mathcal{L} = 0.5 \times \text{CE} + \frac{\alpha}{2 \times (\alpha+\beta)} \cdot \text{Dice} + \frac{\beta}{2 \times (\alpha+\beta)} \cdot \text{X}\), where \(\text{X}\) denotes cl-X-Dice or B-DoU \cite{sun2023boundary}. Parameters \(\alpha\) and \(\beta\) are non-negative numbers; with both set to 0, it reverts to CE loss.
\begin{figure}[t]
\centering
\includegraphics[width=\textwidth]{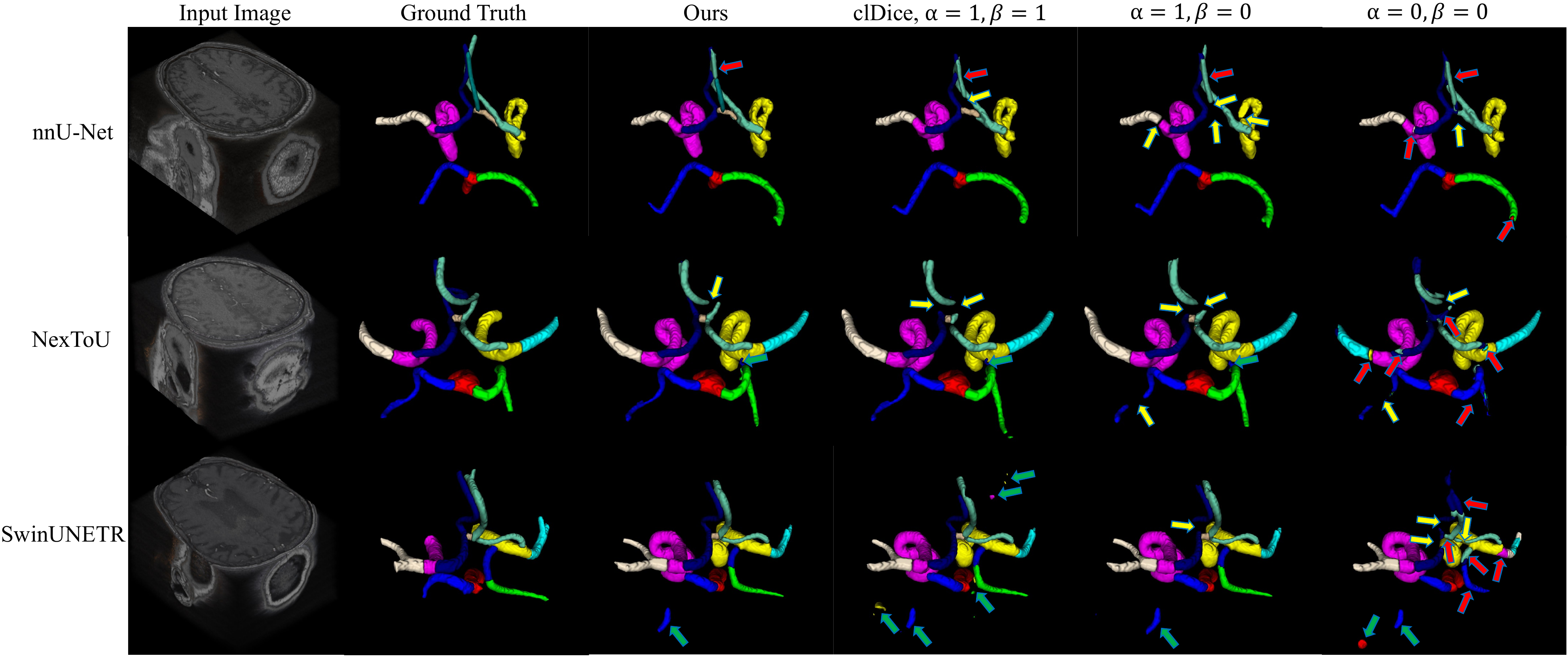}
\caption{Comparative visualization of results on the TopCoW 2023 dataset. Yellow arrows mark areas of segmentation false negatives, green arrows point to false positives, and red arrows identify areas of misclassification.} \label{fig4}
\end{figure}
\subsection{Results}
Our method demonstrates effectiveness in 2D and 3D binary segmentation tasks. Detailed analysis of the DRIVE dataset, as presented in Table~\ref{tab:comprehensive_results_comparison_2D}, demonstrates that the cbDice loss significantly improves overlap metrics. Specifically, it achieves the highest Dice score of 82.5\% and a clDice score of 82.4\%. Additionally, the cbDice variant with \( \beta = 0.5 \) reaches a top NSD of 87.2\%. In parallel, cl-M-Dice performs comparably to B-DoU. Within the Parse 2022 dataset, the cbDice loss is effective in maintaining both topology and geometry in vascular structures, as evidenced by improved segmentation Dice scores, increased NSD scores, and reduced topological errors.

The cbDice loss approach has substantially improved vascular segmentation, especially in precision and topological integrity of small structures. This is particularly significant in the TopCoW 2023 dataset, as shown in Table~\ref{tab:comprehensive_results_comparison_3D}. The metric \(\beta^{\text{err}}\), denoted by '--', is not calculated for multi-class segmentation if not all classes are segmented. Our comparative visualization for the TopCoW, as shown in Figure \ref{fig4}, effectively highlights segmentation errors, with arrows indicating false negatives, false positives, and misclassifications. Our loss function, which incorporates cbDice, contrasts starkly with the standard CE and Dice loss (\( \alpha \)=1, \( \beta \)=0) and the combination of CE, Dice, and clDice loss (\(\text{X=clDice}\), \( \alpha \)=1, \( \beta \)=1). It exhibits superior performance, particularly in multi-class segmentation and enhancing connectivity among different vessel classes. The cbDice approach outperforms conventional Dice and clDice methods in terms of scores, while also reducing false positives and negatives. For example, nnU-Net with cbDice achieves Dice(S) scores as high as 43.38\% at \( \alpha \) and \( \beta \) values of 1 and 2, surpassing clDice's 38.46\%. Similarly, NexToU with cbDice attains a peak Dice(S) of 48.43\% at \( \alpha = 1 \) and \( \beta = 3 \). Overall, the cbDice loss approach represents a significant advancement in accurately segmenting small and complex vascular structures across diverse datasets. For 2D binary segmentation, setting the hyperparameter \( \beta \) to 0.5 or 1 offers advantages. In 3D binary segmentation, \( \beta \) values of 1 or 2 are more effective. For 3D multi-class imbalanced segmentation, a \( \beta \) value of 2 or 3 is recommended. Its effectiveness in maintaining structural integrity and precision underscores the need for tailored \( \beta \) values to optimize performance for specific network architectures and datasets.

\begin{table}[t]
\centering
\caption{Comprehensive comparison of results on the Parse 2022 and TopCoW 2023 datasets. Here, \(\text{L}\) denotes large (non-communicating) arteries, and \(\text{S}\) represents small (communicating) arteries \cite{yang2023benchmarking}.}
\label{tab:comprehensive_results_comparison_3D}
\scriptsize
\begin{tabular*}{\textwidth}{@{\extracolsep{\fill}} c|ccc|cccc|cccccc}
\hline
Network & \multicolumn{3}{c|}{Loss} & \multicolumn{4}{c|}{Parse 2022 (50 epochs)} & \multicolumn{6}{c}{TopCoW 2023 (100 epochs, classes average)} \\ 
\cmidrule(){2-14}
& \(\text{X}\) & \( \alpha \) &  \( \beta \) & Dice & clDice & \(\beta^{\text{err}}\) & NSD & Dice(\(\text{L}\)) & Dice(\(\text{S}\)) & clDice & \(\beta^{\text{err}}\) & NSD(L) & NSD(S) \\
\hline
\multirow{6}{*}{nnU-Net} & -- & 0 & 0 & 82.55 & 68.91 & 331.5 & 78.51 & 81.51 & 7.012 & 90.10 & -- & 89.78 & 14.08 \\
                          & -- & 1 & 0 & 85.05 & 80.11 & 277.7 & 86.04 & 84.03 & 0 & 88.22 & -- & 91.90 & 0 \\
                          & clDice & 1 & 1 & 85.30 & 80.23 & \textbf{263.4} & 86.20 & \textbf{84.21} & 38.46 & 90.72 & -- & 92.45 & 47.85 \\
                          & \cellcolor{lightyellow}cbDice & \cellcolor{lightyellow}1 & \cellcolor{lightyellow}1 & \textbf{85.46} & \textbf{80.76} & 266.1 & \textbf{86.57} & 84.11 & 41.55 & 91.34 & 0.96 & 92.21 & 50.42 \\
                          & \cellcolor{lightyellow}cbDice & \cellcolor{lightyellow}1 & \cellcolor{lightyellow}2 & 84.91 & 80.02 & 275.8 & 85.98 & 84.01 & \textbf{43.38} & \textbf{91.95} & 0.98 & 91.99 & \textbf{54.11} \\
                          & \cellcolor{lightyellow}cbDice & \cellcolor{lightyellow}1 & \cellcolor{lightyellow}3 & 84.97 & 78.98 & 285.6 & 85.26 & 84.18 & 42.68 & 90.63 & \textbf{0.95} & \textbf{92.67} & 51.76 \\
\cline{1-14}
\multirow{6}{*}{SwinUNETR} & -- & 0 & 0 & 78.91 & 58.53 & 508.7 & 70.80 & 61.22 & 0 & 88.64 & -- & 70.64 & 0 \\
                          & -- & 1 & 0 & 81.94 & 69.87 & 496.9 & 78.33 & \textbf{83.19} & 37.32 & 90.18 & 1.40 & 89.55 & 46.36 \\
                          & clDice & 1 & 1 & 82.06 & 70.12 & 499.4 & 78.80 & 83.16 & 34.72 & 90.03 & 1.31 & \textbf{90.73} & 43.65 \\
                          & \cellcolor{lightyellow}cbDice & \cellcolor{lightyellow}1 & \cellcolor{lightyellow}1 & 
                          \textbf{82.19} & \textbf{71.04} & 476.5 & \textbf{79.36} & 82.29 & 37.34 & 90.21 & 1.43 & 89.51 & 46.49 \\
                          & \cellcolor{lightyellow}cbDice & \cellcolor{lightyellow}1 & \cellcolor{lightyellow}2 & 81.88 & 69.96 & \textbf{469.0} & 78.89 & 82.86 & 38.38 & \textbf{90.56} & 1.29 & 90.70 & \textbf{48.37} \\
                          & \cellcolor{lightyellow}cbDice & \cellcolor{lightyellow}1 & \cellcolor{lightyellow}3 & 81.59 & 68.62 & 477.5 & 77.54 & 83.09 & \textbf{38.85} & 90.24 & \textbf{1.27} & 90.67 & 47.52 \\
\cline{1-14}
\multirow{6}{*}{NexToU}   & -- & 0 & 0 & 81.33 & 64.80 & 387.7 & 75.35 & 56.33 & 0 & 89.93 & -- & 67.79 & 0 \\
                          & -- & 1 & 0 & 85.08 & 79.39 & 288.5 & 85.49 & \textbf{84.52} & 42.37 & 90.44 & 0.99 & 92.30 & 51.82 \\
                          & clDice & 1 & 1 & 84.88 & 79.27 & 274.3 & 85.23 & 84.30 & 43.90 & 90.19 & 0.66 & 92.02 & 53.31 \\
                          & \cellcolor{lightyellow}cbDice & \cellcolor{lightyellow}1 & \cellcolor{lightyellow}1 & 85.19 & \textbf{80.07} & 248.95 & \textbf{85.88} & 84.32 & 44.51 & 90.72 & 0.68 & 92.18 & 55.62 \\               
                          & \cellcolor{lightyellow}cbDice & \cellcolor{lightyellow}1 & \cellcolor{lightyellow}2 & \textbf{85.22} & 79.72 & \textbf{216.2} & 85.56 & 84.37 & 47.39 & 90.45 & \textbf{0.65} & \textbf{92.37} & 57.71 \\
                          & \cellcolor{lightyellow}cbDice & \cellcolor{lightyellow}1 & \cellcolor{lightyellow}3 & 85.05 & 79.18 & 289.6 & 85.31 & 84.21 & \textbf{48.43} & \textbf{90.89} & 0.66 & 92.30 & \textbf{58.91} \\
\hline
\end{tabular*}
\end{table}

\section{Conclusion}
In this study, we introduce the cbDice loss, an innovative advancement within the clDice loss framework, further elaborated through a series of cl-X-Dice metrics. These developments specifically cater to the complexities of vascular segmentation in medical imaging. Extensive evaluations across diverse datasets have validated the efficacy of our method in preserving topological integrity, capturing geometric detail, and maintaining balanced diameter representation in vascular segmentation tasks. Comparative analysis against leading models and loss functions have demonstrated the superior capability of the cbDice loss in meeting the intricate demands of vascular segmentation.

\begin{credits}
\subsubsection{\ackname} This work was supported in part by the National Natural Science Foundation of China under Grant 62276081 and 62106113, in part by the National Key Research and Development Program of China under Grant 2021YFC2501202, and in part by the Major Key Project of Peng Cheng Laboratory under Grant PCL2023A09.

\subsubsection{\discintname} The authors have no competing interests to declare that are relevant to the content of this article.
\end{credits}

%
%
%
\bibliographystyle{splncs04}
\bibliography{cbDice_reference}

\newpage
\newpage
\section{Supplementary Material} 
\noindent\textbf{Theoretical Analysis of clDice Variants.} This section delves into the theoretical foundations and geometric sensitivities of cl-X-Dice metrics in vascular segmentation. We introduce three theorems to elucidate the behavior and computation of cl-X-Dice metrics:

\begin{theorem}
\label{theorem:1}
For vertical translations along skeleton lines without deformation, cl-M-Dice coefficient is sensitive to translations of mask $V$ within radius $R$, whereas clDice remains invariant.
\end{theorem}

\begin{proof}
In 2D, cl-M-Dice is defined thus (extendable analogously to 3D):
\begin{equation}
\mathrm{Tprec}(S_{\mathrm{P}}, S_{\mathrm{L}}, V_{\mathrm{L}}) = \frac{|S_{\mathrm{P}} \cap D_{\mathrm{L}}|}{|R_{\mathrm{P}} \cap (U - S_{\mathrm{L}})| + |S_{\mathrm{P}} \cap R_{\mathrm{L}}|}
\end{equation}
\begin{equation}
\mathrm{Tsens}(S_{\mathrm{L}}, S_{\mathrm{P}}, V_{\mathrm{P}}) = \frac{|S_{\mathrm{L}} \cap D_{\mathrm{P}}|}{|R_{\mathrm{L}} \cap (U - S_{\mathrm{P}})| + |S_{\mathrm{L}} \cap R_{\mathrm{P}}|}
\end{equation}
Under vertical translations maintaining constant radius, \( |S_{\mathrm{P}} \cap R_{\mathrm{L}}| \) equals \( |S_{\mathrm{L}} \cap R_{\mathrm{P}}| \). This reduces cl-M-Dice's denominator to \( |R_{\mathrm{P}}| \) (and analogously for \( R_{\mathrm{L}} \)), making its sensitivity dependent solely on the numerator. Hence, cl-M-Dice reacts to spatial displacements of \( V \) within \( R \).
Conversely, clDice, assessing overlap between \( S \) and \( V \), is not influenced by these variations.
\end{proof}
  
\begin{theorem}
\label{theorem:2}
cl-S-Dice, unlike clDice, is sensitive to radius variations at the skeleton under deformation without perpendicular translation. In cases of complete overlap, cl-S-Dice equals clDice with a value of 1.
\end{theorem}

\begin{proof}
In 2D, cl-S-Dice is defined thus (extendable analogously to 3D):
\begin{equation}
\mathrm{Tprec}(S_{\mathrm{P}}, S_{\mathrm{L}}, V_{\mathrm{L}}) = \frac{|R_{\mathrm{P}} \cap V_{\mathrm{L}}|}{|R_{\mathrm{P}}|}, \quad \mathrm{Tsens}(S_{\mathrm{L}}, S_{\mathrm{P}}, V_{\mathrm{P}}) = \frac{|R_{\mathrm{L}} \cap V_{\mathrm{P}}|}{|R_{\mathrm{L}}|}
\end{equation}
For clDice \( \neq 1 \) (partial overlap), changes in radius (\( R_{\mathrm{P}} \), \( R_{\mathrm{L}} \)) affect both \( |R_{\mathrm{P}} \cap V_{\mathrm{L}}| \) and \( |R_{\mathrm{L}} \cap V_{\mathrm{P}}| \). Specifically, with \( S = \{s_i, b^{\text{s}}_j \mid i \in [1, n], j \in [1, m] \} \) and \( R = \{r_i \cdot s_i, b^{\text{s}}_j \mid, i \in [1, n], j \in [1, m] \} \), variances in \( r_i \) at any \( s_i \) modify cl-S-Dice.
When clDice \( = 1 \) (complete overlap), \( |R_{\mathrm{P}} \cap V_{\mathrm{L}}| = |R_{\mathrm{P}}| \) and \( |R_{\mathrm{L}} \cap V_{\mathrm{P}}| = |R_{\mathrm{L}}| \), aligning cl-S-Dice with clDice, highlighting cl-S-Dice's sensitivity to radius changes in other scenarios.
\end{proof}     

\begin{theorem}
\label{theorem:3}
cl-X-Dice enhances geometric sensitivity and compensates for diameter differences while preserving clDice's topological integrity.
\end{theorem}

\begin{proof}
The cl-X-Dice metric, through the incorporation of variables \( Q_{\mathrm{sl}}, Q_{\mathrm{sp}}, Q_{\mathrm{vl}}, \) and \( Q_{\mathrm{vp}} \), offers an advanced sensitivity to geometric alterations, including size and shape variability, while upholding the topological preservation traits of clDice. 
\begin{equation}
\label{eq:Tprec_clXDice}
\mathrm{Tprec}(S_{\mathrm{P}}, S_{\mathrm{L}}, V_{\mathrm{L}}) = \frac{|Q_{\mathrm{sp}} \cap Q_{\mathrm{vl}}|}{|Q_{\mathrm{sp}} \cap Q_{\mathrm{spvp}} \cap (U - S_{\mathrm{L}})| + |Q_{\mathrm{sp}} \cap Q_{\mathrm{slvl}}|} \\
\end{equation}
\begin{equation}
\label{eq:Tsens_clXDice}
\mathrm{Tsens}(S_{\mathrm{L}}, S_{\mathrm{P}}, V_{\mathrm{P}}) = \frac{|Q_{\mathrm{sl}} \cap Q_{\mathrm{vp}}|}{|Q_{\mathrm{sl}} \cap Q_{\mathrm{slvl}} \cap (U - S_{\mathrm{P}})| + |Q_{\mathrm{sl}} \cap Q_{\mathrm{spvp}}|} \\
\end{equation}
Eq.~\ref{eq:Tprec_clXDice} and Eq.~\ref{eq:Tsens_clXDice} represent a balanced approach, maintaining topological accuracy while adapting to geometric variances, thus achieving an equilibrium between topological integrity and geometric precision.
\end{proof}

\begin{table}[!ht]
\centering
\begin{tabular}{@{}llcccccc@{}}
\toprule
& & \multicolumn{3}{c}{CoW Anterior Variants} & \multicolumn{3}{c}{CoW Posterior Variants} \\ 
\cmidrule(lr){3-5} \cmidrule(lr){6-8}
& & \includegraphics[width=1.2cm]{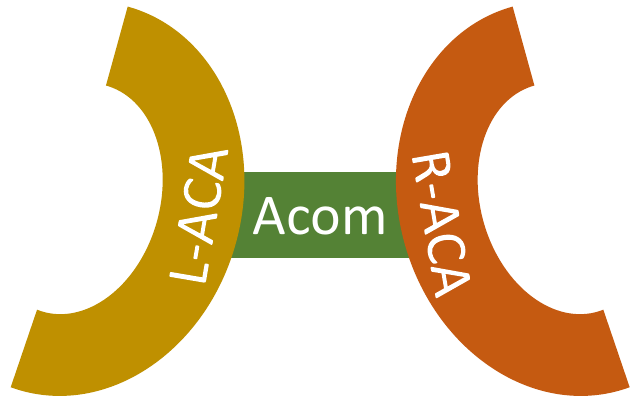} & \includegraphics[width=1cm]{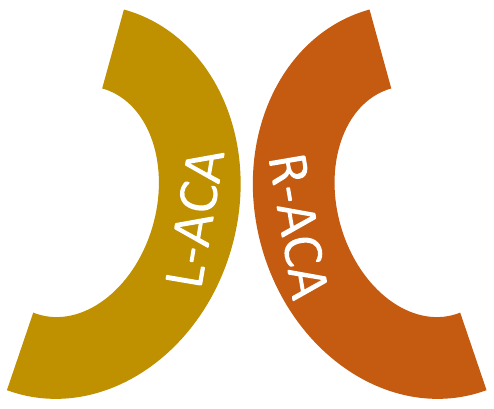} & \includegraphics[width=1.2cm]{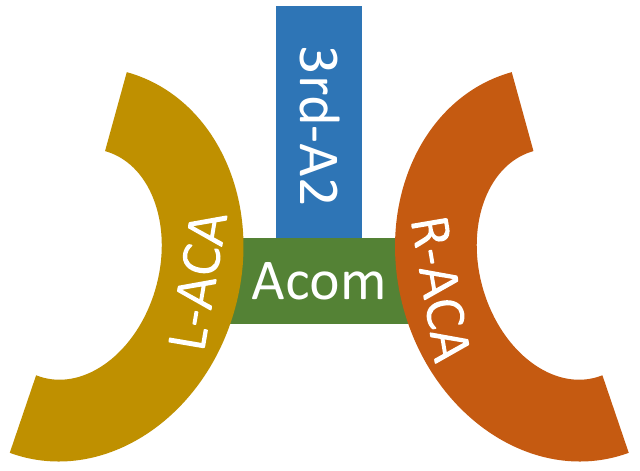} & \includegraphics[width=1cm]{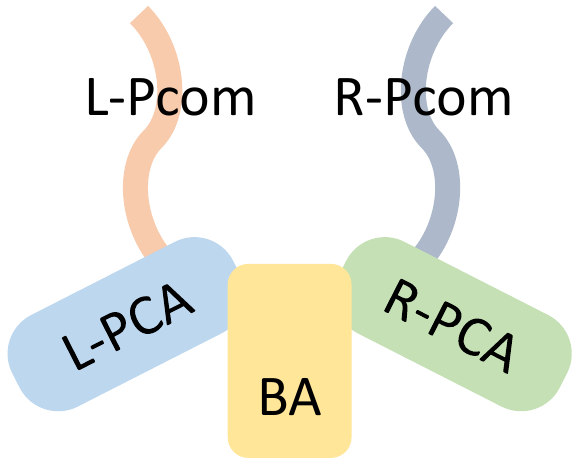} & \includegraphics[width=1cm]{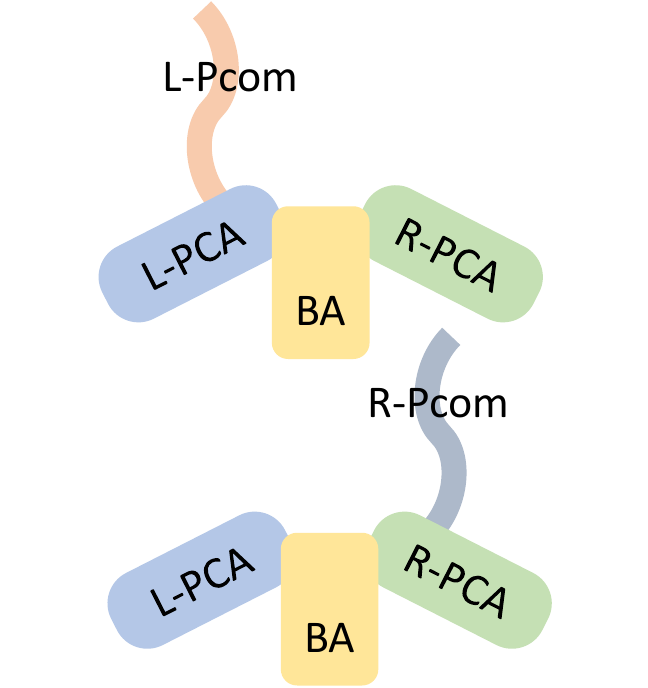} & \includegraphics[width=1cm]{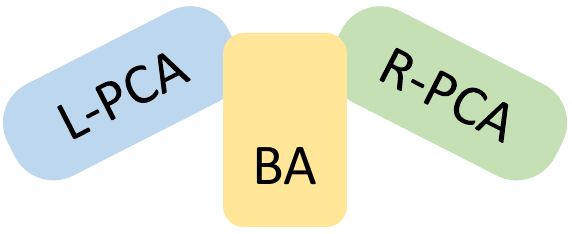} \\
& & (n=11) & (n=2) & (n=5) & (n=4) & (n=6) & (n=8) \\
\midrule
\multirow{3}{*}{nnU-Net} 
  & CE+Dice & 0\% & 0\% & 0\% & 0\% & 17\% & 50\% \\
  & CE+Dice+clDice & 64\% & 50\% & 0\% & 75\% & 67\% & 75\% \\
  & CE+Dice+cbDice & 64\% & 50\% & 40\% & 75\% & 83\% & 75\% \\
\midrule
\multirow{3}{*}{SwinUNETR} 
  & CE+Dice & 36\% & 0\% & 20\% & 100\% & 67\% & 50\% \\
  & CE+Dice+clDice & 45\% & 50\% & 0\% & 75\% & 100\% & 63\% \\
  & CE+Dice+cbDice & 45\% & 50\% & 20\% & 100\% & 100\% & 88\% \\
\midrule
\multirow{3}{*}{NexToU} 
  & CE+Dice        & 55\% & 50\% & 40\% & 50\% & 50\% & 50\% \\
  & CE+Dice+clDice & 73\% & 50\% & 40\% & 100\% & 67\% & 63\% \\
  & CE+Dice+cbDice & 73\% & 50\% & 60\% & 100\% & 67\% & 75\% \\
\bottomrule
\end{tabular}
\caption{CoW variant topology matching performance on the TopCoW 2023.}
\end{table}

\newcolumntype{Y}{>{\centering\arraybackslash}X}
\newcolumntype{C}[1]{>{\centering\arraybackslash}p{#1}}
\begin{table}[!ht]
\centering
\begin{tabularx}{\textwidth}{lC{2cm}Y}
\toprule
Category & Abbreviation & Full Name \\
\midrule
\multirow{9}{*}{\begin{tabular}[c]{@{}l@{}}Non-communicating\\ arteries\end{tabular}} 
  & BA & Basilar Artery \\
  & R-PCA & Right Posterior Cerebral Artery \\
  & L-PCA & Left Posterior Cerebral Artery \\
  & R-ICA & Right Internal Carotid Artery \\
  & R-MCA & Right Middle Cerebral Artery \\
  & L-ICA & Left Internal Carotid Artery \\
  & L-MCA & Left Middle Cerebral Artery \\
  & R-ACA & Right Anterior Cerebral Artery \\
  & L-ACA & Left Anterior Cerebral Artery \\
\midrule
\multirow{4}{*}{\begin{tabular}[c]{@{}l@{}}Communicating\\ arteries\end{tabular}} 
  & R-Pcom & Right Posterior Communicating Artery \\
  & L-Pcom & Left Posterior Communicating Artery \\
  & Acom & Anterior Communicating Artery \\
  & 3rd-A2 & A2 segment of the Anterior Cerebral Artery \\
\bottomrule
\end{tabularx}
\caption{Classification of the Cerebral Arteries in the Circle of Willis}
\end{table}

\end{document}